# Young's two-slits experiment with people

Raoul Nakhmanson[1]

*The two-slits and one-slit experiments with people are performed
and show the existence of wave component of women behavior.*

Two-slits experiment, firstly performed by Thomas Young with light in 1801, played an important role in the history of physics. Originally it had solved an apparent wave-particle dilemma of light (Huygens vs. Newton). The wave concept had won and gained the upper hand in the whole 19$^{th}$ century. But in the beginning of 20$^{th}$ century the parity was restored and "*dilemma*" was changed to "*duality*".

After appearing of wave mechanics (de Broglie-Schrödinger) the interference experiments were successfully spread on massive particles, namely, electrons, neutrons, atoms, and molecules [1].

Let us consider the two-slits Gedankenexperiment in details (Fig.1). Some electrons (photons, neutrons,...) emitted from the source $S$ reach the registration plane $R$ because the screen $SC$ has two slits. In the plane $R$ the electrons are registered as particles, but their spatial distribution looks like an interference of two coherent waves cut out by the screen $SC$ from the initial wave generated by the source $S$. To emphasize is that it is not a many-body effect: The pattern is the same in the case when the source $S$ is so week that no more than one electron flies between $S$ and $R$ in the same time, and we must conclude there is a "self-interference" of electrons. Seemingly electrons (and other particles) present themselves as well as waves. Hence the notion "wave-particle duality".

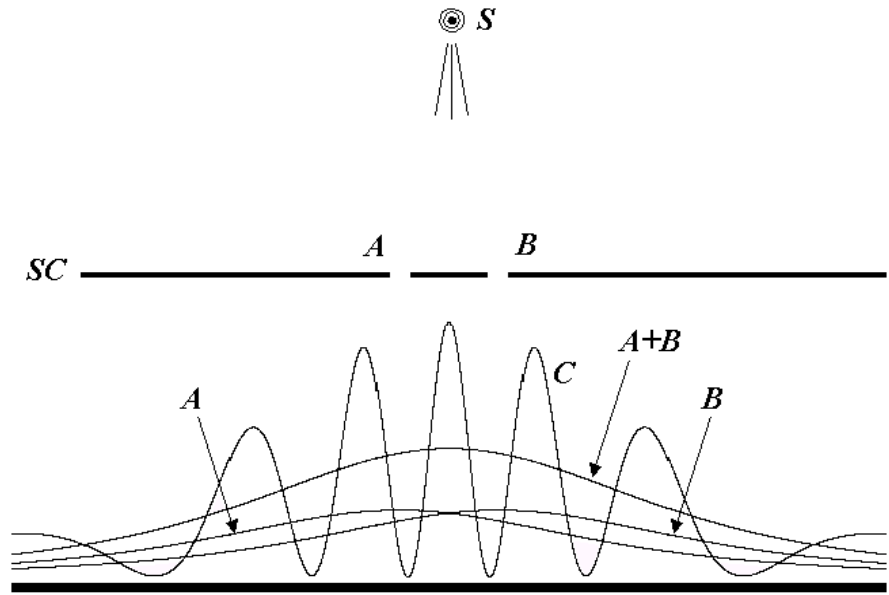

Fig. 1. Two-slits experiment. The distributions $A$, $B$, and $C$ correspond to the cases when the slit $A$ only, the slit $B$ only, and the slits $A$, $B$ both are opened, respectively.

To have the interference pattern we must accept that the "wave of matter" goes through both slits. But if we place two detectors behind the slits, only one of them registers the electron, hence it goes only through one slit. The other slit at the moment seemingly does not play any role and can be temporarily closed. If it is so, the two-slits experiment could be decomposed into two one-slit experiments (curve $A+B$ in Fig. 1) which do not possess the interference being observed. Therefore we have an apparent conflict, and it is an essence of the "interference mystery".

---

[1] nakhmanson@t-online.de  15.02.2004

The good conditions for two-slits experiment are $t < w < \lambda < d < D < s$, where $t$ is a thickness of the screen $SC$, $w$ is a width of the slits, $\lambda$ is the wave length of particles, $d$ is the distance between the slits, $D$ is the distance between the screen and the registration plane $R$, and $s$ is the distance between $S$ and $SC$. The situation is thought as two-dimensional. The energy (and the number of particles) coming to the unit surface of $R$ if, say, the slit $A$ is open and the slit $B$ is closed, is $E_a \sim \cos^2(\varphi_a)*(r_a)^{-1}$, and wave amplitude is $F_a \sim \cos(\varphi_a)*(r_a)^{-1/2} * \exp(-i2\pi r_a/\lambda)$, where $r_a$ is the length of the vector from the slit $A$ to the point $x$ of the registration plane $R$ and $\varphi_a$ is the angle between this vector and normal to $R$. The amplitude of waves from the slit $B$ comes with index $b$ instead of $a$. If the both slits are opened the amplitude $F_a$ and $F_b$ must be added and energy is a squared module of these sum:

$$E \sim |F_a+F_b|^2 \sim E_a+E_b+2(E_a*E_b)^{1/2}*\cos[2\pi(r_a-r_b)/\lambda] . \qquad (1)$$

In Fig. 1 $\lambda = 4$, $d = 14$, and $D = 40$ (relative units).

A real two-slits experiment with "single" electrons was performed in Bologna in 1976. But as Gedankenexperiment it was discussed earlier to emphasize the paradoxical essence of quantum world and to understand its laws. Richard Feynman returned to it in his lectures oft. He said: "We choose to examine a phenomenon which is impossible, absolutely impossible, to explain in any classical way, and which has in it the heart of quantum mechanics. In reality, it contains the only mystery."

The author of this article has his own local-realistic interpretation of quantum mechanics explaining all its paradoxes [2]. I think elementary particles are very complex products of evolution [3] and can receive, work out and spread semantic information. With other words, they have some kind of consciousness ("hidden parameter") and can predict future – the ability which was missed by John Bell.

We know the particles are ruled by the wave laws. The question is: Do they have any consciousness? Several experiments were suggested to prove it [2]. From other side we know that people have their consciousness. The question is: Are they ruled by wave laws? To find it out we must look for interference.

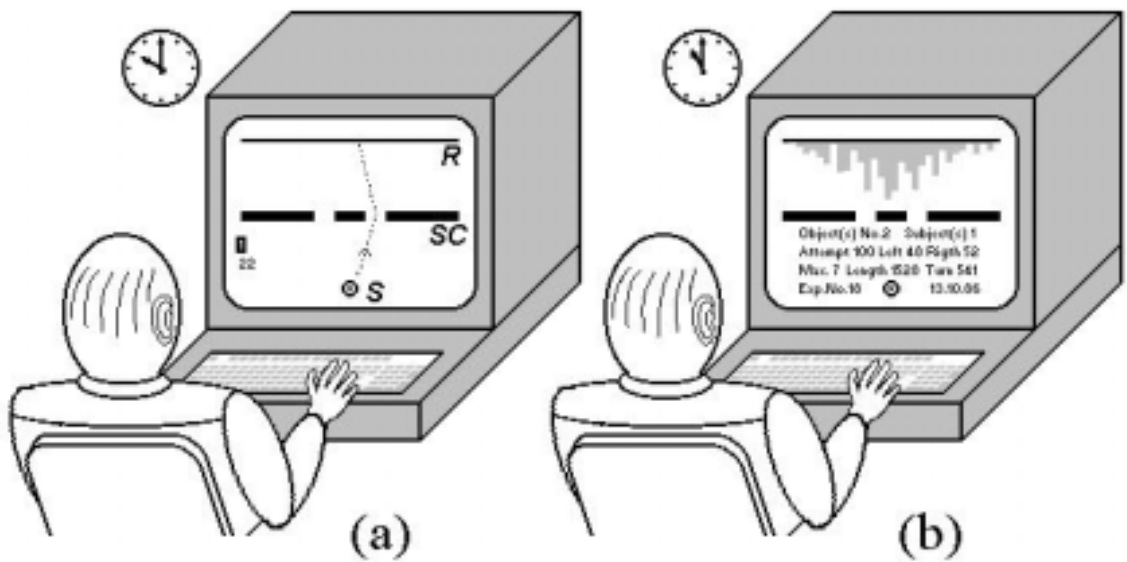

Fig. 2. Computer simulation of two-slits experiment with people.

In 1986-1987 (Novosibirsk) I tried to perform a computer-simulated two-slits experiment with people (Fig. 2(*a*)). The "particle" comes from the source *S* and moves up. The person under test can deflect it left and right, particularly can lead it through one of the slits in the screen *SC* and further to the registration plane *R* . As in case with particles the main deflection can take place near the slits (before, in, and after).

The space between *S* and *R* was defined as a "forest". In the forest there are random distributed hidden "mushrooms" which develop themselves only if the particle touch them. The new mushrooms grow randomly over the forest while their quantity stays constant. The person was motivated to pick the mushrooms. The trajectory of moving object ended in one of 31 channels dividing the plane *R* . After 100 to 200 attempts the distribution of hits over the channels of registration plane *R* was displayed (Fig. 2(*b*)). For comparison the program has also screens with only left or only right slit. Prior to start the experiment the test person had a possibility to make herself "warm" with a sub-program where she can make mushrooms visible. After starting of experiment the mushrooms were new randomized. The test persons were asked not to work with a plan but spontaneous.

To get a stable result in such experiments one needs good statistics. It can be one person having stable characteristics and be tested repeatedly, or many persons having the same characteristics. Both requests correspond to "monochrome" flow and are not easy to realize. Another problem is the density of mushrooms. Because the collision with a mushroom can be regarded as identification of the position of a particle and a possible change of its path, that is, the collapse of a wave function, such attempts were excluded from the statistics. Therefore for statistics it is better to have no mushrooms at all, that however destroys the motivation.

Although of the first results were promising, the problem with statistics was not solved then. The experiments were stopped and are continued only in 2003 (Frankfurt am Main). Programs were improved[2] particularly the number of channels was increased to 63. During the test a calm classical music was played via headphones. Thirty one test persons had performed 63213 attempts (trajectories) in 68 sessions during circa 90 hours. The ages of the test persons were between 6 and 43 years, mainly between 20 and 30 (students).

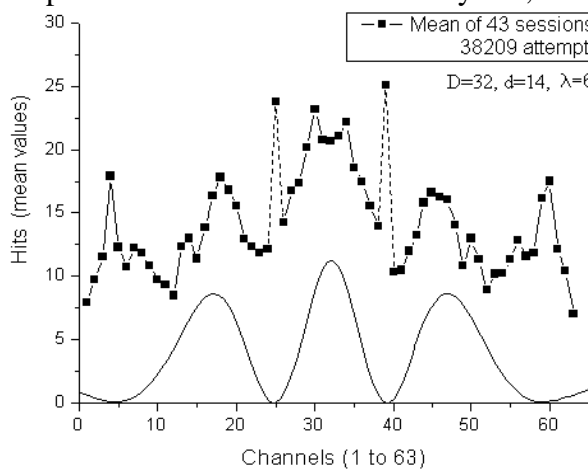 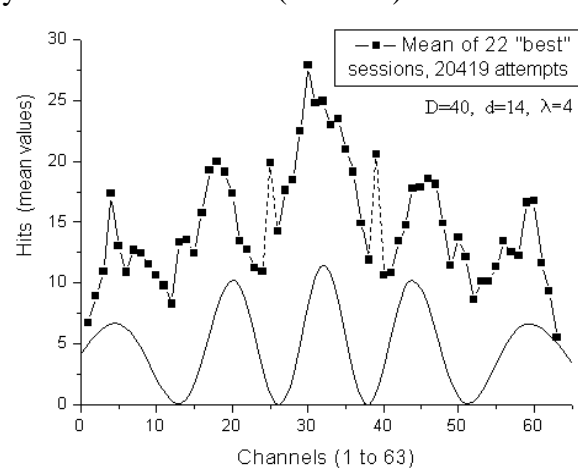

Fig. 3. Means. All 43 sessions are included.    Fig. 4. 22 sessions selected as "wave-like".

Figs. 3 and 4 present the averaged results of the two-slits experiment with people. There were 26 persons (23 women and 3 men) providing 43 sessions with 38209 attempts. The wave component of distribution is evident, and the contrast of the central peaks is more than 35%. Moreover, there are gentle hints at fine structures. The positions of slits correspond to channels No. 25 and No. 39 . One can see lonely peaks there (stroke lines). These are thought artifacts appearing in the moments of relaxing when test persons simple leave the object without control.

---

[2] The last versions of the programs are free and can be sent per e-mail on request.

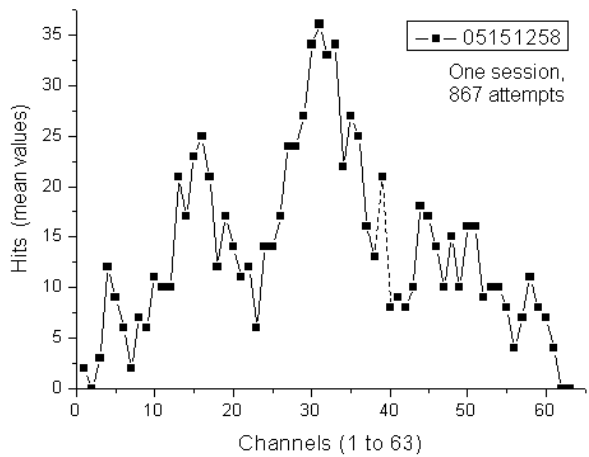 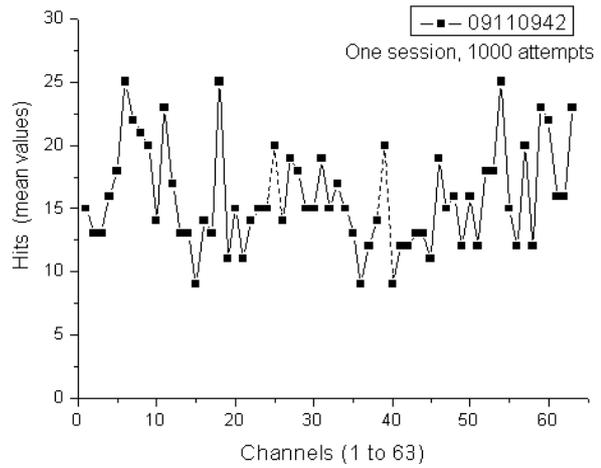

Fig. 5. "Wave-like" session.    Fig. 6. "Homogeneous" session.

It must be emphasized that test persons in the case of Fig. 3 were not selected, all being available were included. The distributions of individual sessions were different, two extreme examples are presented in Fig. 5 and Fig. 6. Therefore the ensemble of sessions of test persons was far from being homogeneous ("monochrome") leading to washout of interference fringes. Nevertheless the results can be regarded as representative: The distribution of the "best half" (22 sessions selected as "wave-like", 20419 attempts) shown in Fig. 4 is not far away from one shown in Fig. 3 and their standard deviations $\sigma$ are homogeneous (Figs. 7 and 8).

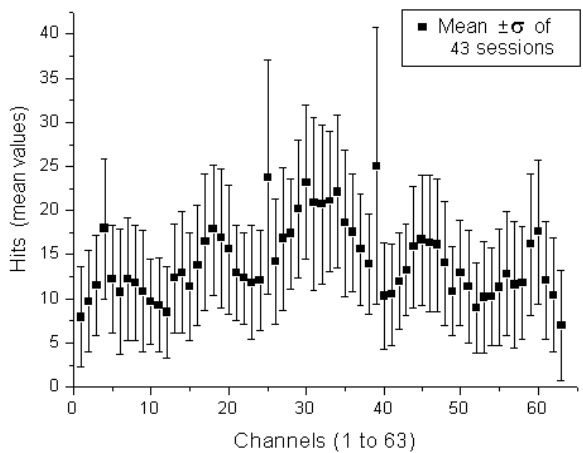 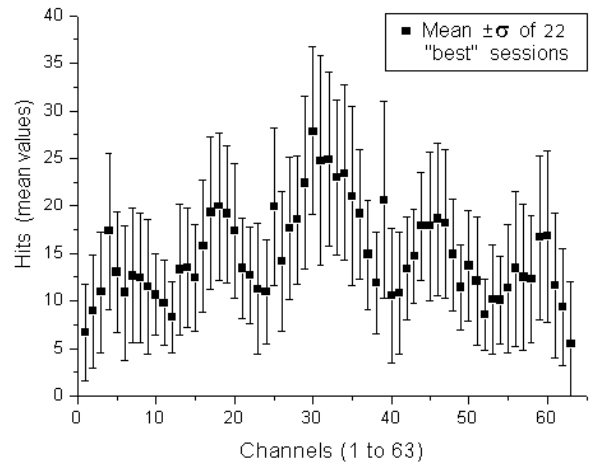

Fig. 7. Standard deviations $\sigma$ of 43-ensemble.    Fig. 8. Standard deviations $\sigma$ of 22-ensemble.

The solid curve in Fig. 3 corresponds to the formula (1) with $\mathbf{D} = 32$, $\mathbf{d} = 14$, and $\boldsymbol{\lambda} = 6$ (relative values). The $\mathbf{d} = 14$ appears as the number of channels between slits (39-25), the ratio $\mathbf{D}/\mathbf{d}$ is known from the geometry of experiment, and $\boldsymbol{\lambda}$ is a free parameter being tuned to fit the positions of peaks. The compromise is found at $\boldsymbol{\lambda} = 6$ but it is far from to be good. The discrepancy can be indebted to one or more of following reasons:

1. Inhomogeneous "non-monochrome" ensemble of sessions of test persons.

2. Computer simulation is not a reality. Particularly the eyes of a test person were farther from plane $\mathbf{R}$ as "eyes" of moving object that can be accepted by the test person as extension of the distance $\mathbf{D}$. The ratio between $\mathbf{d}$ and $\boldsymbol{\lambda}$ seems more solid. This ratio can be found e.g. from the number of minima $\mathbf{N} \approx 2\mathbf{d}/\boldsymbol{\lambda}$ and does not depend on **D**. The distributions shown in Figs. 3 and 4 tend to **N** ≥ 6. If **d** = 14 then λ ≤ 4.7. The solid line in Fig. 4 shows a better agreement if **D** = 40, **d** = 14 and λ = 4.

3. The wave equation of people is not the same as the wave equation of electrons. As it will be seen below it is really so.

The two-slits experiment showed that the wave component of men is vanishing small, particularly they were not included in the "best 22". In 43-ensemble the contribution of men as compared with one of women was negligible (3 men with 3 session and 2880 attempts against 23 women with 40 sessions and 35329 attempts). Therefore the distributions on Figs. 3 and 4 can be regarded as women ones. There was few data of men in two-slits experiment to make some conclusions, but it was clear that women and men can be regarded separately.

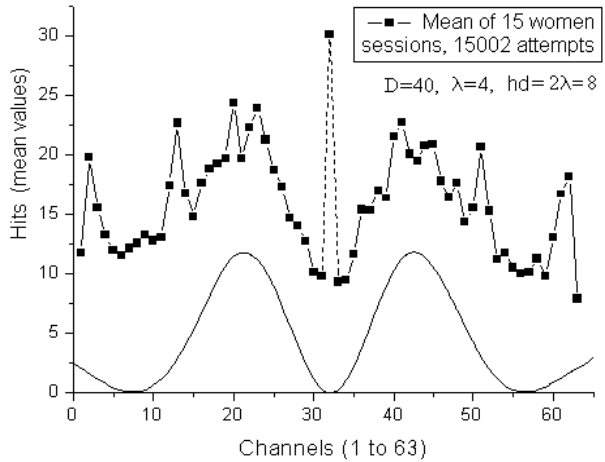
Fig. 9. One slit. Means of women hits.

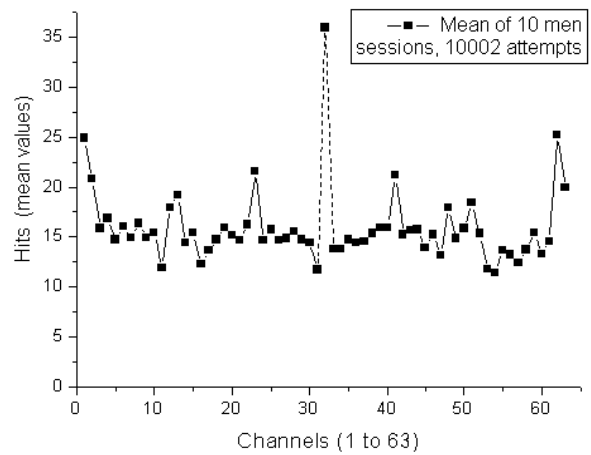
Fig. 10. One slit. Means of men hits.

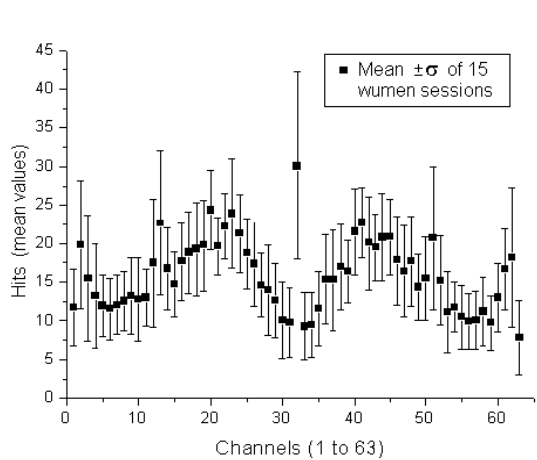
Fig. 11. Standard deviations of women hits.

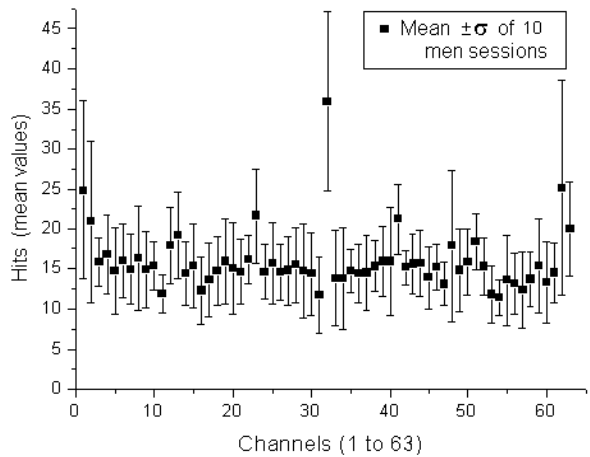
Fig. 12. Standard deviations of men hits.

Such a possibility was used in one-slit experiment with 3 women (15 sessions, 15002 attempts) and 2 men (10 sessions, 10002 attempts). Nobody of these persons took part in the two-slits experiment. Figs. 9 and 11 present the results of women, Figs. 10 and 12 present the results of men. The wave component of distribution of women hits is evident with the contrast more than 40% and hints at fine structures are also seen. On the contrary the distribution of men hits shows only "fine structures". The mean standard deviation averaged over all 63 channels was $\langle\sigma\rangle$ = 5.37 for women and $\langle\sigma\rangle$ = 4.88 for men in spite of the number of men was 1.5 times smaller. That means, the ensemble of men sessions was more homogeneous.

The position of the slit corresponds to the central channel No. 32. One can see a lonely peak there, on Figs. 9 and 10 it is marked by stroke lines. This is thought an artifact appearing in the moments of relaxing when test persons simple leave the object without control. It can be also a "Buridan's ass" reaction on a symmetrical situation.

Whereas the result of two-slits experiment with people is similar to the results obtained with particles (curve ***C*** in Fig. 1, solid curves in Figs. 3 and 4), in the one-slit experiments with people the results differ from ones obtained with particles (curves ***A***, ***B*** in Fig. 1). Especially surprising is distribution of women hits, that is, the deep minimum instead of maximum in the center (Figs. 9 and 11). The one approximation can be obtained if we alter the sign of the interference member in (1):

$$\mathbf{E} \sim \mathbf{E}_a + \mathbf{E}_b - 2(\mathbf{E}_a * \mathbf{E}_b)^{1/2} * \cos[2\pi(\mathbf{r}_a - \mathbf{r}_b)/\lambda] \ . \tag{2}$$

Because we do not have two slits here, the value of "hidden **d**" (**hd**) must be defined as compared with the wave length $\lambda$, for example **hd** = $2\lambda$. The shape of the "theoretical" curve depends then only on the ratio **D**/$\lambda$. The solid curve in Fig. 9 is obtained using (2) with **D** = 40, $\lambda$ = 4, and **hd** = 8, and be additionally scaled to have the same wave amplitude.

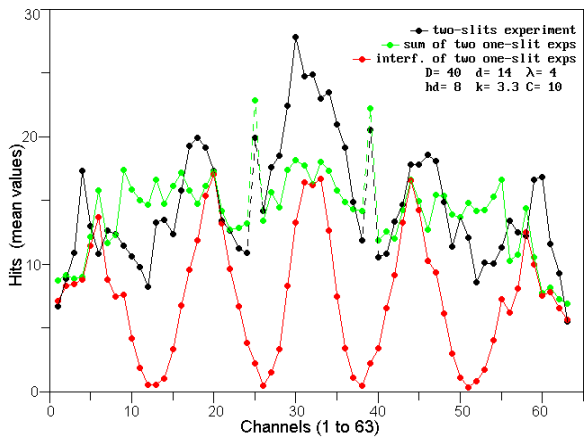 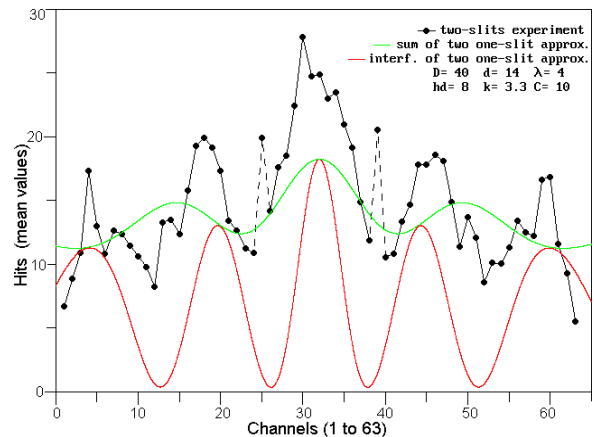

Fig. 13. Two-slits experiment as composed by two one-slit experiments

Fig. 14. Two-slits experiment as composed by two one-slit approximations

In spite of the surprising distribution shown in Fig. 9 it is not in a total contradiction with results of the two-slits experiment. In Fig. 13 the black circles presents once more the experimental two-slits results of "best 22" sessions. The green circles present the simple sum of two experimental one-slit results (scaled to have the same attempt-to-session ratio) displaced in -7 and +7 channels i.e. in two-slits positions, respectively. Therefore the green distribution corresponds to the curve ***A*** + ***B*** in Fig. 1. The red circles in Fig. 13 present the interference of these displaced experimental one-slit results in accordance with the formula (1) at **D** = 40, **d** = 14, and $\lambda$ = 4. To have a better view the amplitude of red distribution is reduced by two.

Fig. 14 is an analog of Fig. 13 but instead of experimental one-slit distribution its approximation (solid line in Fig. 9 displaced in a constant C=+10) was used.

*Acknowledgement:* Many thanks to staff of Institute for Psychology of Goethe University (Frankfurt, Germany) especially to W. Mack and U. Krause for their help in experiments.